\documentstyle[12pt]{article}                    
\newfont{\ffont}{msym10}                        
\newcommand{\beq}{\begin{equation}}             
\newcommand{\eeq}{\end{equation}}               
\newcommand{\bqry}{\begin{eqnarray}}            
\newcommand{\eqry}{\end{eqnarray}}              
\newcommand{\bqryn}{\begin{eqnarray*}}          
\newcommand{\eqryn}{\end{eqnarray*}}            
\newcommand{\preprint}[1]{\begin{table}[t]      
            \begin{flushright}                  
            \begin{large}{#1}\end{large}        
            \end{flushright}                    
            \end{table}}                        
\newcommand{\PD}[2]                             
    {\frac{\partial^{#2}}{\partial #1^{#2}}}    
\begin{document}
\preprint{FERMILAB-PUB-97/265-T \\ LA-UR-97-XXXX}
\title{New Quadratic Mass Relations \\ for Heavy Mesons}
\author{\\ L. Burakovsky\thanks{E-mail: BURAKOV@PION.LANL.GOV. 
Present address: Theoretical Division, MS B285, Los Alamos National 
Laboratory, Los Alamos, NM 87545, USA}, \ 
L.P. Horwitz\thanks{E-mail: HORWITZ@TAUNIVM.TAU.AC.IL. Permanent address: 
School of Physics and Astronomy, Tel-Aviv University, Ramat-Aviv, Israel. 
Also at Department of Physics, Bar-Ilan University, Ramat-Gan, Israel } \
\\  \\ Fermi National Acceleration Laboratory \\ Batavia, IL 60510, USA
\\  \\ and \\  \\ 
T. Goldman\thanks{E-mail: GOLDMAN@T5.LANL.GOV} \
\\  \\  Theoretical Division, MS B283 \\  Los Alamos National Laboratory \\ 
Los Alamos, NM 87545, USA \\}
\date{ }
\maketitle
\begin{abstract}
By assuming the existence of (quasi)-linear Regge trajectories for heavy 
mesons, we derive new quadratic mass relations of non-Gell-Mann--Okubo type, 
$6M^2(q\bar{q})+3M^2(c\bar{c})=8M^2(c\bar{q}),$ $20M^2(q\bar{q})+5M^2(b\bar{
b})=16M^2(b\bar{q}),$ $q=n(=u,d),s$ which show excellent agreement with 
experiment. We also establish the sum rule $M^2(i\bar{i})+M^2(j\bar{j})-2M^2(j
\bar{i})\approx {\rm const}$ for any pair of flavors, $(i,j).$
\end{abstract}
\bigskip
{\it Key words:} flavor symmetry, quark model, heavy mesons, 
Gell-Mann--Okubo, Regge phenomenology

PACS: 11.30.Hv, 11.55.Jy, 12.40.Nn, 12.40.Yx, 14.40.Lb, 14.40.Nd
\bigskip

\section{Introduction}
The generalization of the standard $SU(3)$ Gell-Mann--Okubo mass formula 
\cite{GMO} to higher symmetry groups, e.g., $SU(4)$ and $SU(5),$ became a
natural subject of investigation after the discovery of the fourth and fifth 
quark flavors in the mid-70's \cite{disc}. Attempts have been made in the 
literature to derive such a formula, either quadratic or linear in mass, by a)
using group theoretical methods \cite{Mac,BMO,Boal}, b) generalizing the 
perturbative treatment of $U(3)\times U(3)$ chiral symmetry breaking and the 
corresponding Gell-Mann-Oakes-Renner relation \cite{GOR} to $U(4)\times U(4)$ 
\cite{GLR,MRT}, c) calculating the corrections to the Gell-Mann--Okubo formula
in the charmed-quark sector due to second order $SU(4)$ breaking effects, 
using current algebra techniques \cite{SS}, d) assuming the asymptotic 
realization of $SU(4)$ symmetry in the algebra $[A_\alpha ,A_\beta ]=if_{
\alpha \beta \gamma }V_{\gamma }$ (where $V_{\alpha },A_{\beta }$ are vector 
and axial-vector charges, respectively) \cite{HO}, e) extending the Weinberg 
spectral function sum rules \cite{Wei} to accommodate the higher symmetry 
breaking effects \cite{BGN}, and f) applying alternative methods, such as 
Regge phenomenology \cite{R1,R2}, and the linear mass spectrum for meson 
multiplets\footnote{Here, we speak of linear spectrum over the additive, $I_3$
and $Y,$ multiplet quantum numbers, taking proper account of degeneracy, but do
not (directly) make use of linear Regge trajectories.} \cite{lin,su4}. In the 
following\footnote{Here $\eta _n$ stands for the masses of both isovector and 
isoscalar $n\bar{n}$ states which coincide on a naive quark model level.}, 
$\eta _n,\eta _s,\eta _c,\eta _b,K,D_n,D_s,B_n,B_s,B_c$ stand for the masses 
of the $n\bar{n}$ $(n\equiv u,d),$ $s\bar{s},c\bar{c},b\bar{b},s\bar{n},c\bar{
n},c\bar{s},b\bar{n},b\bar{s},b\bar{c}$ mesons, respectively\footnote{Since 
these designations apply to all spin states, vector mesons will be confusingly
labelled as $\eta $'s. We ask the reader to bear with us in this in the 
interest of minimizing notation.}, unless otherwise specified. The linear mass
relations (here the symbol for the meson represents its mass value)
\bqry
D_n & = & \frac{\eta _n+\eta _c}{2},\;\;\;D_s\;=\;\frac{\eta _s+\eta _c}{2},
 \\  
B_n & = & \frac{\eta _n+\eta _b}{2},\;\;\;B_s\;=\;\frac{\eta _s+\eta _b}{2},\;
\;\;B_c\;=\;\frac{\eta _c+\eta _b}{2}
\eqry  
found in \cite{Boal,MRT}, although perhaps justified for vector mesons,
since a vector meson mass is given approximately by a sum of the corresponding 
constituent quark masses, $$m(i\bar{j})\simeq m(i)+m(j)$$ (in fact, for vector
mesons, the relations (1),(2) hold with an accuracy of up to $\sim 4$\%),
are expected to fail for other meson multiplets, as confirmed by direct
comparison with experiment. Similarly, the quadratic mass relation
\beq
D_s^2-D^2=K^2-\pi ^2
\eeq
obtained in ref. \cite{GLR} by generalizing the $SU(3)$ Gell-Mann-Oakes-Renner 
relation \cite{GOR} to include the $D$ and $D_s$ mesons,
\beq
\frac{\pi ^2}{2n}=\frac{K^2}{n+s}=\frac{D^2}{n+c}=\frac{D_s^2}{s+c},
\eeq
(and therefore $D_s^2-D^2=K^2-\pi ^2\propto (s-n),$ also found in refs. 
\cite{BMO,HO,BGN}), does not agree with experiment. For pseudoscalar mesons,
for example, one has (in GeV$^2)$ 0.388 for the l.h.s. of (3) vs. 0.226 for 
the r.h.s.. For vector mesons, the corresponding quantities are  0.424 vs. 
0.199, with about 100\% discrepancy. The reason that the relation (3) does not
hold is apparently due to the impossibility of perturbative treatment of $U(4)
\times U(4)$ symmetry breaking, as a generalization of that of $U(3)\times U(3
),$ due to very large bare mass of the $c$-quark as compared to those of the 
$u$-, $d$- and $s$-quarks. 

It was concluded in ref. \cite{SS} that second order $SU(4)$ breaking effects
shift the masses of the charmed vector and pseudoscalar mesons, respectively,
upwards and downwards from the values predicted by the quadratric 
Gell-Mann--Okubo type formula $(I$ stands for isospin)
\beq
\frac{1}{2}\left( M^2(n\bar{n},\;I=1)+M^2(n\bar{n},\;I=0)\right) +M^2(c\bar{
c})=2M^2(c\bar{n})
\eeq
obtained by generalizing the standard $SU(3)$ relation \cite{sm}
\beq
\frac{1}{2}\left( M^2(n\bar{n},\;I=1)+M^2(n\bar{n},\;I=0)\right) +M^2(s\bar{
s})=2M^2(s\bar{n}),
\eeq
and hence\footnote{The reason for the appearance of a peculiar combination of 
the pseudoscalar meson squared masses in Eq. (8) is the following assumed 
flavor content of the $\eta $ and $\eta ^{'}$ \cite{DGH},
\bqryn
\eta & \simeq  & 0.58\;(u\bar{u}+d\bar{d})-0.57\;s\bar{s}\;\simeq \;\frac{u
\bar{u}+d\bar{d}-s\bar{s}}{\sqrt{3}}, \\
\eta ^{'} & \simeq  & 0.40\;(u\bar{u}+d\bar{d})+0.82\;s\bar{s}\;\simeq \;
\frac{u\bar{u}+d\bar{d}+2s\bar{s}}{\sqrt{6}},
\eqryn
as explained below in the text.} \cite{SS}
\beq
\frac{1}{2}\left( \rho ^2+\omega ^2\right) +(J/\psi )^2-2D^{\ast 2}\simeq -0.60
\;{\rm GeV}^2,
\eeq
\beq
\frac{\pi ^2}{2}+\frac{\eta ^2}{3}+\frac{\eta ^{'2}}{6}+\eta _c^2-2D^2\simeq 
0.80\;{\rm GeV}^2.
\eeq
In fact, charmed meson masses of all (four) well-established multiplets are 
shifted downwards, and the magnitudes of these shifts seem not to depend on 
the quantum numbers of the corresponding multiplets; indeed, using the 
measured meson masses, one obtains 
\beq
\frac{1}{2}\left( \rho ^2+\omega ^2\right) +(J/\psi )^2-2D^{\ast 2}=2.12\pm 
0.02\;{\rm GeV}^2,
\eeq
\beq
\frac{\pi ^2}{2}+\frac{\eta ^2}{3}+\frac{\eta ^{'2}}{6}+\eta _c^2-2D^2=2.17\pm
0.02\;{\rm GeV}^2,
\eeq
and for and pseudovector and tensor mesons, respectively,
\beq
\frac{1}{2}\left( b_1^2+h_1^2\right) +h_c^2(1P)-2D_1^2=2.14\pm 0.09\;{\rm GeV
}^2,
\eeq
\beq
\frac{1}{2}\left( a_2^2+f_2^2\right) +\chi _{c2}^2(1P)-2D_2^{\ast 2}=2.23\pm 
0.04\;{\rm GeV}^2,
\eeq
so that Eqs. (9)-(12) agree with each other with an accuracy of up to $\sim $
5\%. Similar conclusion seems to hold in the bottom sector where for two 
established vector and tensor multiplets the corresponding shifts
\beq
\frac{1}{2}\left( \rho ^2+\omega ^2\right) +\Upsilon ^2-2B^{\ast 2}=33.38\pm 
0.04\;{\rm GeV}^2,
\eeq
\beq
\frac{1}{2}\left( a_2^2+f_2^2\right) +\chi _{b2}^2(1P)-2B_2^{\ast 2}=35.01\pm
0.28\;{\rm GeV}^2
\eeq
agree with each other to $\sim 4.5$\%.  

With this background of the apparent failure of the standard methods to 
reproduce proper mass relations for higher symmetry group, alternative methods
turn out to be quite successful. Regge phenomenology explored by the present 
authors in refs. \cite{R1,R2} leads to sixth and fourteenth power mass 
relations (for $SU(4)$ and $SU(5)$ meson multiplets, respectively), out of 
which only the former may be tested so far giving an accuracy of up to $\sim $
5\% for all (four) well-established multiplets. The relation  
\beq
12\bar{D}^2=7\eta _0^2+5\eta_c^2,
\eeq
obtained by two of the present authors in ref. \cite{su4} by the application 
of the linear spectrum to $SU(4)$ meson hexadecuplet (here $\bar{D}$ is the 
average mass of the $D_n$ and $D_s$ states which are mass degenerate when 
flavor $SU(4)$ symmetry is broken down to $SU(3)$ by $m(c)\neq m(s)=m(n),$ and
$\eta _0$ is the mass average of the corresponding $SU(3)$ nonet which is also
mass degenerate in this case), holds with a similar accuracy of up to 5\%.   

The purpose of the present paper is to show that, in addition to the 
quite successful higher power relations obtained in \cite{R1,R2}, quadratic 
mass formulas may be also derived. We recall that such formulas have already 
been presented in refs. \cite{R1,R2} by fitting the values of the Regge slopes
of the corresponding meson trajectories; e.g. \cite{R1},   
\beq
8.13\;K^2+4.75\;\eta _c^2=6\left( D_n^2+D_s^2\right) ,
\eeq
with an accuracy of $\sim $ 1\%. Non-integer coefficients in Eq. (16) reflect 
the uncertainty in fitting the values of the Regge slopes. We shall show that 
no such fitting is required to obtain quadratic meson mass relations which, in
contrast to (16), contain integer coefficients (similar to quadratic baryon 
mass relations obtained by the present authors in ref. \cite{bar}), and are 
therefore more suitable for practical use, e.g., to make predictions for the
masses of the states yet to be discovered in experiment. 

\section{Regge phenomenology}
It is well known that the hadrons composed of light $(u,d,s)$ quarks populate 
linear Regge trajectories; i.e., the square of the mass of a state with 
orbital momentum $\ell $ is proportional to $\ell :$ $M^2(\ell )=\ell /\alpha 
^{'}+\;{\rm const,}$ where the slope $\alpha ^{'}$ depends weakly on the 
flavor content of the states lying on the corresponding trajectory,
\beq
\alpha ^{'}_{n\bar{n}}\simeq 0.88\;{\rm GeV}^{-2},\;\;\;   
\alpha ^{'}_{s\bar{n}}\simeq 0.84\;{\rm GeV}^{-2},\;\;\;   
\alpha ^{'}_{s\bar{s}}\simeq 0.80\;{\rm GeV}^{-2}.   
\eeq
In contrast, the data on the properties of Regge trajectories of hadrons 
containing heavy quarks are almost nonexistent at the present time, although
it is established \cite{BB} that the slope of the trajectories decreases with
increasing quark mass (as seen in (17)) in the mass region of the lowest
excitations. This is due to an increasing (with mass) contribution of the 
color Coulomb interaction, leading to a curvature of the trajectory near the
ground state. However, as the analyses show \cite{BB,KS,QR}, in the asymptotic
regime of the highest excitations, the trajectories of both light and heavy
quarkonia are linear and have the same slope $\alpha ^{'}\simeq 0.9$ GeV$^{-2
},$ in agreement with natural expectations from the string model. 

If one assumes the (quasi)-linear form of Regge trajectories for hadrons with 
identical $J^{PC}$ quantum numbers (i.e., belonging to a common multiplet), 
then one has for the states with orbital momentum $\ell $
\bqryn
\ell & = & \alpha ^{'}_{i\bar{i}}m^2_{i\bar{i}}\;+a_{i\bar{i}}(0), \\    
\ell & = & \alpha ^{'}_{j\bar{i}}m^2_{j\bar{i}}\;\!+a_{j\bar{i}}(0), \\    
\ell & = & \alpha ^{'}_{j\bar{j}}m^2_{j\bar{j}}+a_{j\bar{j}}(0).    
\eqryn
Using now the relation among the intercepts \cite{inter,Kai},
\beq
a_{i\bar{i}}(0)+a_{j\bar{j}}(0)=2a_{j\bar{i}}(0),
\eeq
one obtains from the above relations
\beq
\alpha ^{'}_{i\bar{i}}m^2_{i\bar{i}}+\alpha ^{'}_{j\bar{j}}m^2_{j\bar{j}}=
2\alpha ^{'}_{j\bar{i}}m^2_{j\bar{i}}.
\eeq
In order to eliminate the Regge slopes from this formula, we need a relation 
among the slopes. Two such relations have been proposed in the literature,
\beq
\alpha ^{'}_{i\bar{i}}\cdot \alpha ^{'}_{j\bar{j}}=\left( \alpha ^{'}_{j\bar{i
}}\right) ^2,
\eeq
which follows from the factorization of residues of the $t$-channel poles
\cite{first,KY}, and
\beq
\frac{1}{\alpha ^{'}_{i\bar{i}}}+\frac{1}{\alpha ^{'}_{j\bar{j}}}=\frac{2}{
\alpha ^{'}_{j\bar{i}}},
\eeq
based on topological expansion and the $q\bar{q}$-string picture of hadrons
\cite{Kai}. 

For light quarkonia (and small differences in the $\alpha ^{'}$ values), there
is no essential difference between these two relations; viz., for $\alpha ^{'
}_{j\bar{i}}=\alpha ^{'}_{i\bar{i}}/(1+x),$ $x\ll 1,$ Eq. (21) gives $\alpha 
^{'}_{j\bar{j}}=\alpha ^{'}_{i\bar{i}}/(1+2x),$ whereas Eq. (20) gives $\alpha 
^{'}_{j\bar{j}}=\alpha ^{'}_{i\bar{i}}/(1+x)^2\approx \alpha ^{'}/(1+2x),$ i.e,
essentially the same result to order $x^2.$ However, for heavy quarkonia (and
expected large differences from the $\alpha ^{'}$ values for the light 
quarkonia) these relations are incompatible; e.g., for $\alpha ^{'}_{j\bar{i}}=
\alpha ^{'}_{i\bar{i}}/2,$ Eq. (20) will give $\alpha ^{'}_{j\bar{j}}=\alpha ^{
'}_{i\bar{i}}/4,$ whereas Eq. (21) $\alpha ^{'}_{j\bar{j}}=\alpha ^{'}_{i\bar{
i}}/3.$ One has therefore to choose between these relations in order to proceed
further. Here we use Eq. (21), since it is much more consistent with (19) than
is Eq. (20), which we tested by using measured quarkonia masses in Eq. (19). We
shall justify this choice in more detail in a separate publication \cite{prep}.
Here we only wish to show an explicit relation of Eq. (21) to the shifts of the
masses of the charmed and beauty mesons downwards from their Gell-Mann--Okubo
values which are independent of the multiplet quantum numbers, as we have seen
above in Eqs. (9)-(12) and (13),(14).  

Assume, as usually, that both $J^{PC}$ and $(J+1)^{-P,-C}$ states belong
to common Regge trajectory, and two trajectories on which $(\bar{n}n,\;I=0)$ 
and $(\bar{n}n,\;I=1)$ states lie have approximately equal slopes. Then
\bqryn
\alpha ^{'}_{n\bar{n}} & = & \frac{1}{M^2((J+1)^{-P,-C},n\bar{n},I=1)-
M^2(J^{PC},n\bar{n},I=1)} \\
 & \simeq  & \frac{1}{M^2((J+1)^{-P,-C},n\bar{n},I=0)-M^2(J^{PC},n\bar{n},I=
0)} \\
 & \simeq  & 2\Big/ \left( M^2((J+1)^{-P,-C},n\bar{n},I=1)-M^2(J^{PC},n\bar{
n},I=1)\right.  \\ 
 &  & \left. +\;M^2((J+1)^{-P,-C},n\bar{n},I=0)-M^2(J^{PC},n\bar{n},I=0)
\right) , \\
\alpha ^{'}_{c\bar{c}} & = & \frac{1}{M^2((J+1)^{-P,-C},c\bar{c})-M^2(J^{PC},
c\bar{c})}, \\
\alpha ^{'}_{c\bar{n}} & = & \frac{1}{M^2((J+1)^{-P,-C},c\bar{n})-M^2(J^{PC},
c\bar{n})}.
\eqryn
Using now these expressions in Eq. (21) with $i=n,\;j=c,$ one obtains
$$\frac{1}{2}\left( M^2((J+1)^{-P,-C},n\bar{n},I=1)-M^2(J^{PC},n\bar{n},I=1)
\right. $$ $$\left. +\;M^2((J+1)^{-P,-C},n\bar{n},I=0)-M^2(J^{PC},n\bar{n},I=0)
\right) $$ $$+\;M^2((J+1)^{-P,-C},c\bar{c})-M^2(J^{PC},c\bar{c})$$ $$\simeq 2
\left( M^2((J+1)^{-P,-C},c\bar{n})-M^2(J^{PC},c\bar{n})\right) ,$$ or $$\frac{
1}{2}\left( M^2((J+1)^{-P,-C},n\bar{n},I=1)+M^2((J+1)^{-P,-C},n\bar{n},I=0)
\right) $$ $$+\;M^2((J+1)^{-P,-C},c\bar{c})-2M^2((J+1)^{-P,-C},c\bar{n})$$
\beq 
\simeq \frac{1}{2}\left( M^2(J^{PC},n\bar{n},I=1)+M^2(J^{PC},n\bar{n},I=0)
\right) +M^2(J^{PC},c\bar{c})-2M^2(J^{PC},c\bar{n}),
\eeq
which is the sum rule discussed above. Since all trajectories in a given $(i
\bar{j})$ sector have approximately equal slopes, the numerical value of the 
difference of the squared masses in Eq. (22) does not depend on the quantum 
numbers $J^{PC},$ as we have seen in Eqs. (9)-(12). By replacing the $c$-quark
by the $b$-quark, a similar sum rule may be derived for the bottom sector 
(Eqs. (13),(14)). Moreover, by repeating the above analysis for arbitrary pair
of flavors, $(i,j),$ one may easily establish the general sum rule
$$M^2(i\bar{i})+M^2(j\bar{j})-2M^2(j\bar{i})\approx {\rm const,}$$ where $M^2(
n\bar{n})$ is given in Eq. (24) below. 

We note that the sum rule (22) cannot be obtained by starting from Eq. (20). 
As easily seen, Eq. (20) would lead to, e.g.,
$$\frac{1}{2}\left( a_2^2-\rho ^2+f_2^2-\omega ^2\right) \left( \chi _{c2}^2(
1P)-(J/\psi )^2\right) \simeq \left( D_2^{\ast 2}-D^{\ast 2}\right) ^2,$$
which gives 3.3 GeV$^2$ on the l.h.s. vs. 4.0 GeV$^2$ on the r.h.s., with an
accuracy of $\sim $ 20\% which is much worse than the accuracy of Eqs. (9)-(12)
and (13),(14). This fact should be considered as the best experimental evidence
for additivity of the inverse Regge slopes, Eq. (21), and against factorization
of the slopes, Eq. (20).

No standard Gell-Mann--Okubo type quadratic mass formula is compatible with 
Eqs. (19),(21), except for the standard $SU(3)$ Gell-Mann--Okubo formula 
itself. By standard Gell-Mann--Okubo type one we mean a quadratic mass 
relation the sums of coefficients on both sides of which coincide (e.g., Eq. 
(15) where $12=7+5).$ Indeed, as follows from (19),(21),
$$\alpha ^{'}_{i\bar{i}}m^2_{i\bar{i}}+\alpha ^{'}_{j\bar{j}}m^2_{j\bar{j}}=
\frac{4\alpha ^{'}_{i\bar{i}}\alpha ^{'}_{j\bar{j}}}{\alpha ^{'}_{i\bar{i}}+
\alpha ^{'}_{j\bar{j}}}m^2_{j\bar{i}}.$$ 
Equating the sums of coefficients on both sides of this relations gives
$$\left( \alpha ^{'}_{i\bar{i}}-\alpha ^{'}_{j\bar{j}}\right) ^2=0,\;\;{\rm 
i.e.,}\;\;\alpha ^{'}_{i\bar{i}}=\alpha ^{'}_{j\bar{j}},$$ and this holds
(approximately) only in the $SU(3)$ sector (for $i=n,\;j=s.)$ Thus, proper 
generalization of the standard quadratic Gell-Mann--Okubo formula to higher 
symmetry groups must be of non-Gell-Mann--Okubo type.  

It is also clear from the comparison of the relations (6), and (19) with $i=n,
j=s:$
\beq
\eta _n^2+\eta _s^2=2K^2,
\eeq
that
\beq
\eta _n^2=\frac{1}{2}\left( M^2(n\bar{n},\;I=1)+M^2(n\bar{n},\;I=0)\right) .
\eeq
This relation explains the appearance of a peculiar combination of the 
pseudoscalar meson masses in Eq. (8). Indeed, the pure ``non-strange'' and
``strange'' pseudoscalar isoscalar states which may be constructed from the
physical $\eta $ and $\eta ^{'}$ states should have the masses
\bqryn  
\eta ^2(n\bar{n}) & = & \frac{2}{3}\;\!\eta ^2\;+\;\frac{1}{3}\;\!\eta ^{'2},
 \\
\eta ^2(s\bar{s}) & = & \frac{1}{3}\;\!\eta ^2\;+\;\frac{2}{3}\;\!\eta ^{'2},
\eqryn
respectively, according to their flavor content, as described in Footnote 4.
Therefore, in view of (24),
\bqryn
\eta _n^2 & = & \frac{1}{2}\left( \pi ^2+\eta ^2(n\bar{n})\right) \;=\;
\frac{\pi ^2}{2}\;+\;\frac{\eta ^2}{3}\;+\;\frac{\eta ^{'2}}{6}, \\
\eta _s^2 & = & \eta ^2(s\bar{s})\;=\;\frac{\eta ^2}{3}\;+\;\frac{2\eta ^{'2}}{
3}.
\eqryn

Theoretical basis for such quadratic meson mass relations of 
non-Gell-Mann--Okubo type with integer coefficients
has been established by Bal\'{a}zs in ref. \cite{Bal} where, by using the dual 
topological unitarization approach to confinement region of QCD which is based
on analyticity and generalized ladder-graph dynamics and takes into account
the effect of planar sea-quark loops, it was shown that
\beq
\alpha ^{'}_{c\bar{c}}=\frac{\alpha ^{'}_{n\bar{n}}}{N_1},\;\;\;\alpha ^{'}_{b
\bar{b}}=\frac{\alpha ^{'}_{n\bar{n}}}{N_2},\;\;\;N_1,N_2\;{\rm are\;integer,}
\eeq
and $N_2>N_1>1$ (i.e., $\alpha ^{'}_{b\bar{b}}<\alpha ^{'}_{c\bar{c}}<\alpha 
^{'}_{n\bar{n}}).$ The approach of ref. \cite{Bal} does not however specify 
the numerical values of $N_1$ and $N_2.$ It was suggested by Bal\'{a}zs that
\beq
N_1=3,\;\;\;N_2=9.
\eeq
Also, in ref. \cite{Bal2}, Bal\'{a}zs has suggested 
\bqryn
\alpha ^{'}_{c\bar{c}} & = & \frac{\alpha ^{'}_{n\bar{n}}}{3},\;\;\;
\alpha ^{'}_{c\bar{n}}\;=\;\frac{\alpha ^{'}_{n\bar{n}}}{2}, \\
\alpha ^{'}_{b\bar{b}} & = & \frac{\alpha ^{'}_{n\bar{n}}}{9},\;\;\;
\alpha ^{'}_{b\bar{n}}\;=\;\frac{\alpha ^{'}_{\bar{n}}}{5},
\eqryn
so that $$\frac{2}{\alpha ^{'}_{c\bar{n}}}=\frac{1}{\alpha ^{'}_{c\bar{c}}}+
\frac{1}{\alpha ^{'}_{n\bar{n}}}=\frac{4}{\alpha ^{'}_{n\bar{n}}},$$ and 
$$\frac{2}{\alpha ^{'}_{b\bar{n}}}=\frac{1}{\alpha ^{'}_{b\bar{b}}}+\frac{1}{
\alpha ^{'}_{n\bar{n}}}=\frac{10}{\alpha ^{'}_{n\bar{n}}},$$ confirming (21).

We now show that these values are motivated by the ratio of the constituent
quark masses, and therefore may be perhaps justified for vector mesons. Indeed,
by viewing a vector meson as $m(i\bar{j})=m(i)+m(j),$ and solving Eq. (21) by
introducing $x\equiv \alpha ^{'}_{i\bar{i}}/\alpha ^{'}_{j\bar{j}},$ as 
follows:
\beq
\alpha ^{'}_{j\bar{i}}=\frac{\alpha ^{'}_{i\bar{i}}}{(1+x)/2},\;\;\;
\alpha ^{'}_{j\bar{j}}=\frac{\alpha ^{'}_{i\bar{i}}}{x},
\eeq
Eq. (19) may be rewritten as 
$$4m^2(i)+\frac{4m^2(j)}{x}=\frac{4(m(i)+m(j))^2}{1+x},$$ leading to
\beq
x=\frac{m(j)}{m(i)}.
\eeq  
For $i=s,\;j=c\;(b),$ $m(i)\simeq 0.5$ GeV, $m(j)\simeq 1.5\;(4.5)$ GeV, and
$x\simeq 3\;(9),$ which are the values suggested by Bal\'{a}sz which enter Eq.
(23) (in the approximation $\alpha ^{'}_{s\bar{s}}\approx \alpha ^{'}_{n\bar{
n}}).$ However, even for vector mesons, the formula obtained from 
(19),(21),(26),
\beq
3\tilde{\rho }^2+(J/\psi )^2=3D^{\ast 2},\;\;\;\tilde{\rho }\equiv \frac{
\rho ^2+\omega ^2}{2},
\eeq
gives\footnote{A corresponding formula in the bottom sector, $$45\tilde{\rho }
^2+5\Upsilon ^2=18B^{\ast 2},$$ holds with an accuracy of $\sim $ 7.5\%.} 11.4 
GeV$^2$ on the l.h.s. vs. 12.1 GeV$^2$ on the r.h.s., with an accuracy of 
$\sim 6$\%. This accuracy is not bad by itself but much worse than that of a
new mass relation for vector mesons we suggest in the next Section. The reason 
for this is the physically incorrect assumption of a meson mass being a sum of
the corresponding constituent quark masses. It is well known that in order to 
reproduce meson spectroscopy correctly, one has to include hyperfine 
(spin-spin, spin-orbit and tensor) interaction, in addition to a sum of the 
constituent quark masses, even for vector mesons \cite{BG}.

In this paper we suggest the values for $N_1$ and $N_2$ which are different 
from those of Eq. (26); namely, 
\beq
N_1=2,\;\;\;N_2=4.
\eeq
The value $N_1=2,$ although seems to be a guess consistent with (25), is quite
natural (in contrast to $N_1=3$ of Bal\'{a}zs), since it leads to
\beq
\alpha ^{'}_{c\bar{c}}=\frac{\alpha ^{'}_{n\bar{n}}}{2}\simeq 0.45\;{\rm GeV}^{
-2},
\eeq
in agreement with $\alpha ^{'}_{c\bar{c}}\simeq 0.5\;{\rm GeV}^{-2}$ which has
long been discussed in the literature \cite{Kai,CN,Inami,Kob}. As we show
below, mass relation obtained on the basis of $N_1=2$ is in excellent agreement
with experiment. The value $N_2=4$ cannot be predicted by any known (at least,
to the authors) theoretical approach, but may be anticipated on the basis of 
simple phenomenological arguments which we present below.

\section{New quadratic mass relations} 
Let us start with quadratic mass relation for the $SU(4)$ multiplet built of 
the $u$-, $d$-, $s$-, and $c$-quarks. This relation follows from (19),(21)
with $i=n,\;j=c,$ and (21) with $N_1=2:$
\beq
6\;\eta _n^2+3\;\eta _c^2=8\;D_n^2.
\eeq
In order to test this relation, we calculate the value of $D_n,$ as given by 
(32), using the measured values of $\eta _n$ and $\eta _c,$ and compare it with
experiment. Our results are shown in Table I. The combination $(\pi ^2/2+\eta 
^2/3+\eta ^{'2}/6)$ has been used for $\eta _n^2$ in the case of pseudoscalar 
mesons, as explained above. One sees excellent agreement with experiment.

In the approximation $\alpha ^{'}_{s\bar{s}}\approx \alpha ^{'}_{n\bar{n}},$ 
another mass relation may be written down (which is obtained by replacing the 
$s$-quark for the non-strange quarks in Eq. (32)),
\beq
6\;\eta _s^2+3\;\eta _c^2=8\;D_s^2.
\eeq
This relation is tested in Table II, again by comparing predictions for $D_s$ 
given by (33) with experiment. Now $(\eta ^2/3+2\eta ^{'2}/3)$ has been used 
for $\eta _s^2$ in the case of pseudoscalar mesons. Again one sees very good
agreement with experiment. The reason for poorer agreement with experiment in
the case of the $1^{+-}$ multiplet may be that the $D_{s1}$ is an axial-vector 
meson, not a pseudovector one, or the mixture of both states. If the $D_{s1}$ 
does belong to the $1^{++}$ multiplet, calculation using Eq. (33) will give 
$D_{s1}=2525\pm 2$ MeV, with the accuracy of 0.4 \%. Let us note that in this 
case the accuracy of Eq. (32), as applied to predict the mass of the $D_1,$ 
will not change since both the $a_1$ and $b_1$ mesons have approximately equal
mass of $\sim 1230$ MeV \cite{data}. 

In order to determine the value of $N_2,$ we use the following phenomenological
arguments. First, as we have already remarked in ref. \cite{R2}, the difference
in the constituent quark masses is the only reason for the different slopes of
the corresponding trajectories. Indeed, if one considers sub-$SU(3)$ symmetry 
which incorporates the $u$-, $d$- and $c$-quarks (with the $c$-quark playing 
the same role as $s$-quark in a real world), one has to obtain the same 
quadratic mass relation in this sub-$SU(3)$ sector as the standard 
Gell-Mann--Okubo one in the sub-$SU(3)$ sector which incorporates the $u$-, 
$d$- and $s$-quarks, 
\beq
m^2=a+bZ+c\left[ \;\frac{Z^2}{4}-I(I+1)\right] ,
\eeq
since charm $(C)$ is now playing the role of strangeness, and the 
``supercharge'' $Z=B+C$ is playing the role of hypercharge. However, the actual
mass relation in the $(u,d,c)$ sector, Eq. (32), differs considerably from the 
standard Gell-Mann--Okubo formula, as applied to this sector, $$\eta _n^2+\eta
_c^2=2D_n^2.$$ Since the only discrepancy between the two sub-$SU(3)$ sectors 
mentioned above not taken into account in the formula (34) is the difference 
in the constituent $s$- and $c$-quark masses, this difference is solely 
responsible for the different slopes of the corresponding trajectories, and 
leads to non-Gell-Mann--Okubo type mass relation (32). This difference is also 
responsible for peculiar numerical coefficients of Eq. (32), viz., 6, 3 and 8. 

It is quite natural to assume that the ratio of the slopes in the $i\bar{i}$ 
and $j\bar{j}$ sectors is solely determined by the {\it ratio} of the 
corresponding constituent quark masses, i.e., 
\beq
\frac{\alpha ^{'}_{i\bar{i}}}{\alpha ^{'}_{j\bar{j}}}=F\left(
\frac{m(j)}{m(i)}\right) ,  
\eeq
where $F(x),\;x\equiv m(j)/m(i),$ is some unknown function.\footnote{An example
of such a function, $F(x)=x,$ is given by Eqs. (27),(28), where $\frac{\alpha 
^{'}_{i\bar{i}}}{\alpha ^{'}_{j\bar{j}}}=\frac{m(j)}{m(i)}.$} Then, in 
particular, $$\frac{\alpha ^{'}_{s\bar{s}}}{\alpha ^{'}_{c\bar{c}}}=F\left( 
\frac{m(c)}{m(s)}\right) \simeq F\left( \frac{1.5\;{\rm GeV}}{0.5\;{\rm GeV}}
\right) =F(3).$$ Since $$\frac{\alpha ^{'}_{c\bar{c}}}{\alpha ^{'}_{b\bar{b}}}
=F\left( \frac{m(b)}{m(c)}\right) \simeq F\left( \frac{4.5\;{\rm GeV}}{1.5\;
{\rm GeV}}\right) =F(3)\simeq \frac{\alpha ^{'}_{s\bar{s}}}{\alpha ^{'}_{c
\bar{c}}},$$ and the coefficients of a mass relation are solely determined by 
the ratio of the constituent quark masses, as discussed above, we conclude 
that the mass relation for mesons composed of the $c$- and $b$-quarks should 
have the same coefficients as the mass relation for mesons composed of the 
$s$- and $c$-quarks, respectively (Eq. (33)), i.e.,
\beq
6\;\eta _c^2+3\;\eta _b^2=8\;B_c^2.
\eeq
This relation will now enable one to obtain the value of $N_2.$ It follows 
from (19),(21) with $i=c,\;j=b,$ and (25) with $N_1=2$ that
\beq
\frac{\eta _c^2}{2}+\frac{\eta _b^2}{N_2}=\frac{2\;B_c^2}{1+N_2/2}.
\eeq
By comparing Eqs. (36) and (37), one finds
\beq
N_2=4.
\eeq
Once $N_2$ is known, one can easily derive mass relations for the $SU(4)$ 
multiplet built of the $u$-, $d$-, $s$-, and $b$-quarks, in a way which is 
completely analogous to that for the derivation of Eqs. (32) and (33) above. 
These are 
\beq
20\;\eta _n^2+5\;\eta _b^2=16\;B_n^2,
\eeq
\beq
20\;\eta _s^2+5\;\eta _b^2=16\;B_s^2,
\eeq
and tested in Tables III and IV, respectively, again by comparing their
predictions for $B_n$ and $B_s$ with experiment. So far, firm comparison is 
only possible for vector mesons. As for the tensor multiplet, we consider the 
states $B_J^\ast (5732)$ and $B_{sJ}^\ast (5850)$ discovered recently, whose 
quantum numbers are uncertain, as tensor mesons, since the former has the 
dominant decay modes $B^\ast \pi$ and $B\pi $ \cite{data}, and the latter lies
in the proper mass interval.

Finally, we can use Eq. (36) to predict the masses of the $(b\bar{c})$ states
not measured so far. In Table V these predictions are compared with rough
quark model-motivated estimate $B_c\simeq (\eta _b+\eta _c)/2,$ for vector and
tensor mesons. \\

\begin{center}
\begin{tabular}{|c|c|c|c|} \hline
$J^{PC}$ & Eq. (28), MeV & Ref. [34], MeV & Accuracy, \% \\ \hline
 $0^{-+}$ & $1877\pm 1$ & $1867\pm 3$ & 0.5 \\ \hline 
 $1^{--}$ &     2012    & $2008\pm 2$ & 0.2 \\ \hline 
 $1^{+-}$ & $2397\pm 6$ & $2422\pm 2$ & 1.0 \\ \hline 
 $2^{++}$ & $2450\pm 1$ & $2459\pm 3$ & 0.4 \\ \hline 
\end{tabular}
\end{center}
{\bf Table I.} Comparison of predictions for the masses of the 
$(c\bar{n})$-mesons given by Eq. (32) with the measured masses provided by the
Particle Data Group \cite{data}, for four well-established meson multiplets.
Electromagnetic corrections are included as uncertainties in the mass values.

\begin{center}
\begin{tabular}{|c|c|c|c|} \hline
$J^{PC}$ & Eq. (29), MeV & Ref. [34], MeV & Accuracy, \% \\ \hline
 $0^{-+}$ & $1965.5\pm 1$ & $1968.5\pm 0.5$ & 0.1 \\ \hline 
 $1^{--}$ &     2092      &  $2112\pm 0.5$  & 0.9 \\ \hline 
 $1^{+-}$ &  $2468\pm 8$  &       2535      & 2.6 \\ \hline 
 $2^{++}$ &  $2547\pm 2$  & $2573.5\pm 1.5$ & 1.0 \\ \hline 
\end{tabular}
\end{center}
{\bf Table II.} Comparison of predictions for the masses of the 
$(c\bar{s})$-mesons given by Eq. (33) with the measured masses provided by the
Particle Data Group \cite{data}, for four well-established meson multiplets.

\begin{center}
\begin{tabular}{|c|c|c|c|} \hline
$J^{PC}$ & Eq. (35), MeV & Ref. [34], MeV & Accuracy, \% \\ \hline
 $1^{--}$ &     5359    & $5325\pm 2$  & 0.6 \\ \hline 
 $2^{++}$ & $5728\pm 1$ & $5698\pm 12$ & 0.5 \\ \hline 
\end{tabular}
\end{center}
{\bf Table III.} Comparison of predictions for the masses of the 
$(b\bar{n})$-mesons given by Eq. (39) with the measured masses provided 
by the Particle Data Group \cite{data}, for vector and tensor mesons.

\begin{center}
\begin{tabular}{|c|c|c|c|} \hline
$J^{PC}$ & Eq. (36), MeV & Ref. [34], MeV & Accuracy, \% \\ \hline
 $1^{--}$ &     5410    &  $5416\pm 3$ & 0.1 \\ \hline 
 $2^{++}$ & $5798\pm 2$ & $5853\pm 15$ & 0.9 \\ \hline 
\end{tabular}
\end{center}
{\bf Table IV.} Comparison of predictions for the masses of the 
$(b\bar{s})$-mesons given by Eq. (40) with the measured masses provided 
by the Particle Data Group \cite{data}, for vector and tensor mesons.

\begin{center}
\begin{tabular}{|c|c|c|c|} \hline
$J^{PC}$ & Eq. (32), MeV & $(\eta _b+\eta _c)/2,$ MeV & Discrepancy, \%
 \\ \hline
 $1^{--}$ & 6384 & 6278.5 & 1.6 \\ \hline 
 $2^{++}$ & 6807 & 6734.5 & 1.1 \\ \hline 
\end{tabular}
\end{center}
{\bf Table V.} Comparison of predictions for the masses of the $(b\bar{c
})$-mesons given by Eq. (36) with rough estimate $(\eta _b+\eta _c)/2,$
for vector and tensor mesons.

\section{Concluding remarks}
In this paper we have further explored Regge phenomenology for heavy hadrons
initiated in our previous publications \cite{R1,R2}. We have obtained new
non-Gell-Mann--Okubo quadratic meson mass relations which show excellent
agreement with experiment (as seen in Tables I-IV, the accuracy of these
relations for pseudoscalar, vector and tensor mesons does not exceed 1\%). 
They are
\bqryn
6\;\eta _n^2+3\;\eta _c^2 & = & 8\;D_n^2, \\
6\;\eta _s^2+3\;\eta _c^2 & = & 8\;D_s^2, \\
6\;\eta _c^2+3\;\eta _b^2 & = & 8\;B_c^2, \\
20\;\eta _n^2+5\;\eta _b^2 & = & 16\;B_n^2, \\
20\;\eta _s^2+5\;\eta _b^2 & = & 16\;B_s^2.
\eqryn
We have shown that the sum rules (9)-(12), and (13),(14) are easily explained
in the framework discussed in this paper.   

The values $N_1=2$ and $N_2=4$ that we have suggested predict the following 
values for the Regge slopes:
\bqryn
\alpha ^{'}_{c\bar{n}}\;\simeq \;\alpha ^{'}_{c\bar{s}} & = & \frac{\alpha ^{'
}_{n\bar{n}}}{1.5}\;\simeq \;0.60\;{\rm GeV}^{-2}, \\
\alpha ^{'}_{c\bar{c}} & = & \frac{\alpha ^{'}_{n\bar{n}}}{2}\;\simeq \;0.45\;
{\rm GeV}^{-2}, \\
\alpha ^{'}_{b\bar{n}}\;\simeq \;\alpha ^{'}_{b\bar{s}} & = & \frac{\alpha ^{'
}_{n\bar{n}}}{2.5}\;\simeq \;0.36\;{\rm GeV}^{-2}, \\
\alpha ^{'}_{b\bar{c}} & = & \frac{\alpha ^{'}_{n\bar{n}}}{3}\;\simeq \;0.30\;
{\rm GeV}^{-2}, \\
\alpha ^{'}_{b\bar{b}} & = & \frac{\alpha ^{'}_{n\bar{n}}}{4}\;\simeq \;0.225\;
{\rm GeV}^{-2}.
\eqryn
It is interesting to compare these values with experiment. Since no more than
one state is known to lie on each of the heavy meson trajectories, we again, as
in Section 2, use the assumption that both $J^{PC}$ and $(J+1)^{-P,-C}$ states
belong to common trajectory, and calculate the slope by the formula
$$\alpha ^{'}_{i\bar{i}}=\frac{1}{M^2((J+1)^{-P,-C},i\bar{i})-M^2(J^{PC},i\bar{
i})}.$$
Our results are shown in Table VI (we assume that radially excited states 
2 $^3S_1$ and 2 $^3P_2$ lie on daughter trajectory which is parallel to the 
leading one to which 1 $^3S_1$ and 1 $^3P_2$ belong).
	 
It is seen that the slopes increase as the mass of the state is increased. For 
$c\bar{c}$ and $b\bar{b}$ states, the highest calculated values are not far 
from the values predicted in the present paper. We may conclude, therefore, 
that the trajectories are $not$ linear but rather have curvatures in the 
region of lower spin, and the values for the slopes predicted in this paper are
achieved in the region of higher spin. Similar curvature is also a feature of 
the pion trajectory.\footnote{If one tries, apart from its Goldstone nature, 
to fit the pion to the {\it linear} trajectory on which the $b_1(1231)$ and 
$\pi _2(1670)$ lie $[\ell =0.80\;M^2(\ell )-0.20],$ extrapolation down to 
$\ell =0$ gives $m(\pi )\simeq 0.5$ GeV, much higher than the physical value 
$m(\pi )=0.138$ GeV, which means that the pion trajectory has curvature near 
$\ell =0.$} Since quadratic relations that we have obtained in the paper are 
very accurate even for lower spin states, we conclude that the additivity of 
the inverse slopes (which is the basis of these relations) is a universal 
feature of Regge trajectories which does not depend on spin and holds in the 
curvature region as well.

It is very interesting to determine the actual form of the function $F(x)$ 
discussed in this paper. Since
\bqryn
N_1 & = & \frac{\alpha ^{'}_{s\bar{s}}}{\alpha ^{'}_{c\bar{c}}}\;\equiv \;F
\left( \frac{m(c)}{m(s)}\right) \;\simeq \;F(3)\;=\;2, \\
N_2 & = & \frac{\alpha ^{'}_{s\bar{s}}}{\alpha ^{'}_{b\bar{b}}}\;\equiv \;F
\left( \frac{m(b)}{m(s)}\right) \;\simeq \;F(9)\;=\;4, \\
\eqryn
one may fit $F$ as $$F(x)=\frac{x}{3}+1,$$ i.e., $$\frac{\alpha ^{'}_{i\bar{
i}}}{\alpha ^{'}_{j\bar{j}}}\simeq \frac{1}{3}\frac{m(j)}{m(i)}+1.$$ 
Theoretical (or phenomenological) models for the form of this ratio of the 
slopes $F(x)$ are called for. \\ 

\begin{center}
\begin{tabular}{|c|c|c|c|c|c|} \hline
$j\bar{i}$ & $1\;^1S_0-1\;^1P_1$ & $1\;^3S_1-1\;^3P_2$ & $1\;^3P_0-1\;^3D_1$ &
$2\;^3S_1-2\;^3P_2$ & Present paper   \\ \hline
$c\bar{n}$ & 0.418 & 0.496 &       &       & 0.60   \\ \hline 
$c\bar{s}$ & 0.392 & 0.462 &       &       & 0.60   \\ \hline 
$c\bar{c}$ & 0.281 & 0.327 & 0.392 &       & 0.45   \\ \hline 
$b\bar{n}$ &       & 0.243 &       &       & 0.36   \\ \hline 
$b\bar{s}$ &       & 0.203 &       &       & 0.36   \\ \hline 
$b\bar{b}$ &       & 0.114 &       & 0.201 & 0.225  \\ \hline 
\end{tabular}
\end{center}
{\bf Table VI.} The slopes of the heavy meson trajectories given by Eq. (37),
in which pairs of states shown in the Table are used (in GeV$^{-2}),$ vs.
predictions of the present paper.


\bigskip
\bigskip
\begin{thebibliography}{9}
\bibitem{GMO} S. Okubo, Prog. Theor. Phys. {\bf 27} (1962) 949, {\bf 28} (1962)
24 \\ M. Gell-Mann and Y. Ne'eman, {\it The Eightfold Way,} (Benjamin, NY, 
1964)
\bibitem{disc} J.J. Aubert {\it et al.,} Phys. Rev. Lett. {\bf 33} (1974) 1404
 \\ J.-E. Augustin {\it et al.,} Phys. Rev. Lett. {\bf 33} (1974) 1406 \\
C. Bacci {\it et al.,} Phys. Rev. Lett. {\bf 33} (1974) 1408 \\ 
G.S. Abrams {\it et al.,} Phys. Rev. Lett. {\bf 33} (1974) 1453 \\
S.W. Herb {\it et al.,} Phys. Rev. Lett. {\bf 39} (1977) 252
\bibitem{Mac} A.J. Macfarlane, J. Phys. G {\bf 1} (1975) 601
\bibitem{BMO} S. Borchardt, V.S. Mathur and S. Okubo, Phys. Rev. Lett. 
{\bf 34} (1975) 38 \\
V.S. Mathur, S. Okubo and S. Borchardt, Phys. Rev. D {\bf 11} (1976) 2572 \\ 
H. Hayashi {\it et al.,} Ann. Phys. {\bf 101} (1976) 394
\bibitem{Boal} D.B. Lichtenberg, Nuovo Cim. Lett. {\bf 13} (1975) 346 \\
A. Kazi, G. Kramer and D.H. Schiller, Acta Phys. Austr. {\bf 45} (1976) 65 \\
D.H. Boal and R. Torgerson, Phys. Rev. D {\bf 15} (1977) 327 \\ 
D.H. Boal, Phys. Rev. D {\bf 18} (1978) 3446 \\ S. Iwao, Nuovo Cim. Lett. 
{\bf 20} (1977) 347, {\bf 21} (1978) 239, 245, {\bf 22} (1978) 192 \\
S.A. Rashid, Indian J. Pure Appl. Phys. {\bf 19} (1981) 172
\bibitem{GOR} M. Gell-Mann, R.J. Oakes and B. Renner, Phys. Rev. {\bf 175}
(1968) 2195
\bibitem{GLR} M.K. Gaillard, B.W. Lee and J.L. Rosner, Rev. Mod. Phys. {\bf 47}
(1975) 277
\bibitem{MRT} C. Montonen, M. Roos and N. T\"{o}rnqvist, Nuovo Cim. Lett. {\bf
12} (1975) 627
\bibitem{SS} R. Simard and M. Suzuki, Phys. Rev. D {\bf 12} (1975) 2002
\bibitem{HO} H.L. Hallock and S. Oneda, Phys. Rev. D {\bf 18} (1978) 841,
{\bf 19} (1979) 347, {\bf 20} (1979) 2932 \\
M. Majewski and W. Tybor, Acta Phys. Pol. B {\bf 9} (1978) 177
\bibitem{Wei} S. Weinberg, Phys. Rev. Lett. {\bf 18} (1967) 507
\bibitem{BGN} B. Bagchi, V.P. Gautam and A. Nandi, Nuovo Cim. Lett. {\bf 24}
(1979) 175 \\ V.P. Gautam, B. Bagchi and A. Nandy, J. Phys. G {\bf 5} (1979) 
885
\bibitem{R1} L. Burakovsky, T. Goldman and L.P. Horwitz, New mass relations
for heavy quarkonia, hep-ph/9704440; to be published
\bibitem{R2} L. Burakovsky, T. Goldman and L.P. Horwitz, New mass relation
for meson 25-plet, hep-ph/9704432; to be published
\bibitem{lin} L. Burakovsky and L.P. Horwitz, Nucl. Phys. A {\bf 609} (1996)
585, {\bf 614} (1997) 373; Found. Phys. Lett. {\bf 9} (1996) 561
\bibitem{su4} L. Burakovsky and L.P. Horwitz, Gell-Mann--Okubo mass formula
for $SU(4)$ meson hexadecuplet [hep-ph/9608300], Found. Phys. Lett, {\it in 
press}
\bibitem{sm} L. Burakovsky, Found. Phys. {\bf 27} (1997) 315
\bibitem{DGH} J.F. Donoghue, E. Golowich and B.R. Holstein, {\it Dynamics of
the Standard Model,} (Cambridge University Press, 1996), p. 205
\bibitem{bar} L. Burakovsky, T. Goldman and L.P. Horwitz, New quadratic baryon
mass relations, hep-ph/9706464; to be published
\bibitem{BB} J.-L. Basdevant and S. Boukraa, Z. Phys. C {\bf 28} (1985) 413; 
Ann. Phys. (Paris) {\bf 10} (1985) 475
\bibitem{KS} J.S. Kang and H. Schnitzer, Phys. Rev. D {\bf 12} (1975) 841
\bibitem{QR} C. Quigg and J.L. Rosner, Phys. Rep. {\bf 56} (1979) 167
\bibitem{rec} K.P. Das and R.C. Hwa, Phys. Lett. B {\bf 68} (1977) 459 \\
T. Tashiro {\it et al.,} Z. Phys. C {\bf 35} (1987) 21 \\
A.B. Batunin, B.B. Kiselyev and A.K. Likhoded, Yad. Phys. {\bf 49} (1989) 554
\bibitem{fra} A. Capella and J. van Tran Thanh, Z. Phys. C {\bf 10} (1981) 249;
Phys. Lett. B {\bf 114} (1982) 450 \\
A.B. Kaidalov, Phys. Lett. B {\bf 116} (1982) 459 \\
A.B. Kaidalov and K.A. Ter-Martirosyan, Sov. J. Nucl. Phys. {\bf 39} (1984) 979
 \\ A.B. Kaidalov and O.I. Piskounova, Z. Phys. C {\bf 30} (1986) 145 \\
G.I. Lykasov and M.N. Sergeenko, Z. Phys. C {\bf 52} (1991) 635
\bibitem{inter} K. Kawarabayashi, S. Kitakado and H. Yabuki, Phys. Lett. B {\bf
28} (1969) 432 \\ R.C. Brower, J. Ellis, M.G. Schmidt and J.H. Weis, Nucl. 
Phys. B {\bf 128} (1977) 175 \\ V.V. Dixit and L.A. Balazs, Phys. Rev. D {\bf
20} (1979) 816
\bibitem{Kai} A.B. Kaidalov, Z. Phys. C {\bf 12} (1982) 63
\bibitem{first} J. Pasupathy, Phys. Rev. Lett. {\bf 37} (1976) 1336 \\ 
K. Igi, Phys. Lett. B {\bf 66} (1977) 276; Phys. Rev. D {\bf 16} (1977) 196 
\bibitem{KY} M. Kuroda and B.-L. Young, Phys. Rev. D {\bf 16} (1977) 204
\bibitem{prep} L. Burakovsky, T. Goldman and L.P. Horwitz, in preparation
\bibitem{Bal} L.A.P. Bal\'{a}zs, Phys. Rev. D {\bf 26} (1982) 1671
\bibitem{Bal2} L.A.P. Bal\'{a}zs, Phys. Lett. B {\bf 99} (1981) 481
\bibitem{BG} L. Burakovsky and T. Goldman, Towards resolution of the enigmas of
$P$-wave meson spectroscopy, hep-ph/9703271; to be published
\bibitem{CN} N.-P. Chang and C.A. Nelson, Phys. Rev. Lett. {\bf 35} (1975) 
1492; Phys. Rev. D {\bf 13} (1976) 1345
\bibitem{Inami} T. Inami, Phys. Lett. B {\bf 56} (1975) 291
\bibitem{Kob} N.A. Kobylinsky, Acta Phys. Pol. B {\bf 10} (1979) 433
\bibitem{data} Particle Data Group, Phys. Rev. D {\bf 54} (1996) 1
\end{thebibliography}
\end{document}